\documentclass{article}

\usepackage[preprint]{neurips_2026}
\usepackage[utf8]{inputenc}
\usepackage[T1]{fontenc}
\usepackage{hyperref}
\usepackage{url}
\usepackage{booktabs}
\usepackage{amsfonts}
\usepackage{amssymb}
\usepackage{nicefrac}
\usepackage{microtype}
\usepackage{xcolor}
\usepackage{graphicx}
\usepackage{amsmath}
\usepackage{xspace}
\usepackage{multirow}
\usepackage{placeins}

\newif\ifstubmode
\stubmodefalse

\definecolor{stubgray}{gray}{0.32}
\definecolor{stubblue}{rgb}{0.10,0.30,0.60}

\ifstubmode
  \newcommand{\stub}[2]{%
    \par\noindent{\color{stubgray}\small\textbf{\color{stubblue}[#1w]}~\itshape #2\par}%
  }
  \newcommand{\cites}[1]{{\color{stubblue}\small\,$\rightarrow$\,#1}}
\else
  \newcommand{\stub}[2]{}
  \newcommand{\cites}[1]{}
\fi

\title{Red-Teaming the Reward: A Verifier and Benchmark\\
for Triton Kernel Optimization Agents}

\author{%
  Pouya Shiri \& Seyed Abbas Hosseini \\
  Turing Enterprises Inc. \\
  \texttt{\{pouya.s, abbas.hosseini\}@turing.com} \\
}

\begin{document}

\maketitle

\begin{abstract}
Benchmarks for coding agents increasingly reward execution speed, but reported speedups may arise from invalid shortcuts or altered program behavior rather than genuine optimization. This problem is especially acute in GPU kernel optimization, where reduced numerical precision or CUDA Graph replay can produce apparent speedups without improving the requested kernel. We present a verifier for Triton kernel optimization, designed from observed frontier-agent behavior, failures in existing verifiers, and adversarial shortcuts. To evaluate it, we introduce KernelQuest: 100 Triton optimization tasks spanning four levels and 21 task families, each with a certified solution that defines an achievable performance target. Six frontier agents attempted each task three times, yielding 1,800 runs. Our verifier rejected 813 runs before their speedups were credited, including 228 that met their targets. In an independent audit, an expert identified 25\% of audited submissions as invalid; three published verifiers and a stopwatch baseline accepted almost all of them, whereas ours accepted almost none. Our verifier accepted all 100 certified solutions. Under adversarial testing, 26 completed runs relied on an exploit, and none was rewarded. Together, these results provide evidence that our verifier is robust to current frontier-agent behaviors, while underscoring that verifier design must evolve as agents become increasingly capable of exploiting reward signals. 
\end{abstract}

\section{Introduction}
\label{sec:intro}

Coding agents are increasingly evaluated not only on whether their code is functionally correct, but also on how efficiently it executes. SWE-bench-style tasks primarily emphasize functional correctness \citep{jimenez2024swebench}, whereas a growing body of benchmarks and training methods additionally rewards runtime performance. SWE-Perf and SWE-Efficiency score agent-generated patches based on the speedups they achieve in real-world repositories \citep{he2025sweperf,ma2025swefficiency}, while GSO evaluates optimization against expert-authored commits used as performance references \citep{shetty2025gso}. Coding agents are also increasingly trained end to end using executable performance rewards \citep{luo2025deepswe}. Performance has therefore become an optimization objective alongside functional correctness. This introduces a central verification challenge: an agent may improve the measured performance score without actually optimizing the program in the intended way.

GPU kernel optimization is one of the clearest examples of this shift. A kernel determines how computation is mapped  onto the GPU and how data moves through memory hierarchy. Therefore, two functionally equivalent kernels can differ substantially in execution time. These performance differences make execution speedup a useful signal for both evaluation and training coding agents. Kernel generation already has dedicated benchmarks \citep{ouyang2025kernelbench} and training approaches that directly optimize agents using execution-performance rewards \citep{li2025cudal1,baronio2025kevin}. However, reward score is typically scored as the ratio between two runtimes measured on the same machine: the runtime of the reference implementation divided by that of the agent-generated implementation. Writing a faster kernel increases this ratio, but so can using lower precision than the reference or capturing the GPU operations in a CUDA graph and replaying them. All three can lead to substantial measured speedups, but only the first performs the optimization requested by the task.

This risk increases as agents become increasingly capable of satisfying evaluation targets without completing
the intended task. Reward models can degrade under optimization pressure
\citep{gao2023overoptimization}, frontier reasoning models can learn to satisfy the checker
rather than the task \citep{baker2025monitoring}, and similar behavior has been observed in
other settings where one program evaluates another, creating opportunities for agents to exploit the evaluator rather than satisfy the intended objective
\citep{skalse2022rewardhacking,helff2026gamingverifiers}. A verifier that relies only on the measured
speedup ratio cannot distinguish among these strategies because each can produce the same score.
The reliability of performance benchmarks therefore depends critically on the verifier. Yet existing benchmarks typically report the scores achieved by agents without systematically verifying whether the submitted optimizations are valid or whether they satisfy the intended task.

To address these limitations, we developed a verifier for kernel optimization in Triton, a language used by much of the recent work in GPU kernel optimization \citep{tillet2019triton}. The proposed verifier was designed using evidence from three sources: observations of frontier agents solving existing public kernel optimization tasks, analysis of the available verifiers used to evaluate those tasks, and adversarial agents explicitly rewarded for maximizing the performance reward while performing as little of the intended task as possible.

A verifier that has not been thoroughly stress-tested provides limited evidence of robustness. To test our proposed verifier across diverse real-world problems, we designed KernelQuest. KernelQuest tasks are expert designed around real-world workloads and span individual operators, fused operator chains, complete
published models, and large transformer models. Each task includes a certified solution whose measured speed defines the target. This exposes false rejections and prevents verifier robustness from being achieved trivially.

Our contributions are \footnote{Visit \url{https://labs.turing.com/benchmarks/kernelquest} for more information on the latest version of KernelQuest benchmark.}:
\begin{itemize}
  \setlength{\itemsep}{0pt}
  \item \textbf{A verifier for kernel optimization evaluated by speedup measurement}, developed from
    observed frontier-agent behavior and from invalid submissions accepted by three published
verifiers plus a stopwatch baseline. The verifier seperately checks for four aspects of a
    submission that can otherwise be misrepresented.
  \item \textbf{KernelQuest}, a benchmark of 100 diverse Triton kernel optimization tasks,
    each paired with a certified solution whose measured performance defines the target. KernelQuest is designed to stress test the verifier reliability rather than to rank agents.
  \item \textbf{A behavioral study of frontier agents under a measured speedup reward},
    covering six agents, three attempts per task, and 1,800 runs. Rejected runs are categorized by the behavior exhibited by the submission rather than by the error it produces.
  \item \textbf{An adversarial audit of the fixed verifier}, in which an adverserial agent was given the KernelQuest tasks and explicitly instructed to maximize the performance reward while performing as little of the intended computational work as possible. The adverserial agent found no exploit that bypassed the verifier.
  \item \textbf{An independent manual audit conducted without access to the verifier}, conducted by an expert without access to the verifier, and based solely on the task instructions.
\end{itemize}
\section{Related work}
\label{sec:related}

Existing verifiers for kernel optimization tasks typically combine correctness testing with runtime measurement. Their main difference lies in how rigorously they validate a submission before accepting its measured performance. KernelBench compares a submission with the reference on random inputs
within a numerical tolerance, and then measures the runtime of both implementations using GPU events \citep{ouyang2025kernelbench}. Because the KernelBench verification only checks the output and measured runtime, it does not capture how the speedup was achieved. Therefore, a submission
that replays a captured CUDA Graph can receive the same speedup score as one that genuinely implements a faster kernel. KernelBench-Verified addresses two weaknesses in this evaluation protocol. It enables
tensor-core arithmetic in the baseline to avoid an artificially slow reference implementation, and it evaluates
correctness across four hidden input distributions rather than a single narrow distribution
that agents had learned to exploit \citep{zhang2026kernelbenchverified}. These changes
strengthen correctness testing, but numerical agreement alone cannot distinguish a genuinely optimized kernel from a submission that achieves a lower measured runtime by capturing GPU operations in a CUDA Graph and replaying them during evaluation. CUDA-L1
synchronizes all GPU work queues before timing ends, after finding that 82 of its 250 generated kernels left work outside the measured interval \citep{li2025cudal1}. This ensures that pending GPU work is included in the runtime measurement, but it does not verify that the measured computation was performed by the kernel written for the task. A simple wall-clock timer with a loose output tolerance provides neither guarantee: work may occur outside the timed interval or bypass the intended kernel entirely.

Other work addresses this problem through complementary approaches, but does not directly verify the runtime behavior of the submitted program. Exploits against KernelBench have produced apparent speedups above $50\times$. Robust-kbench responds by screening kernels with a LLM-based verifiers to screen submissions \citep{lange2025robustkbench}. TritonRL similarly combines rule-based checks with an LLM-based judge to inspect that evaluates whether the submitted code implements the required computation
\citep{woo2025tritonrl}. Both approaches therefore rely in part on an LLM's interpretation of the submitted source. RealisticTritonBench reduces much of the attack surface by evaluating tasks reconstructed from merged pull requests using end-to-end serving latency
\citep{huang2026realistictritonbench}. However, with average speedups near $1.0\times$, the contribution of an individual kernel is difficult to isolate. 

The broader literature shows that this is not unique to kernel optimization. Specification
gaming is well documented in reinforcement learning and learning from human feedback, where agents exploit gaps between the intended objective and the reward signal used to evaluate it \citep{skalse2022rewardhacking,helff2026gamingverifiers}. Unpinned environment state has also
long been known to bias systems benchmarks \citep{mytkowicz2009wrongdata}, and audits of
agentic benchmarks find widespread solution leakage and weak verification
\citep{zhu2025agenticbenchmarks}. These studies expose flaws in benchmark creation, but do not
measure which invalid submissions a verifier would actually accept. Meanwhile, generated
kernels are increasingly being deployed into production serving systems
\citep{xing2026flashinferbench,li2025tritonbench}, while training methods increasingly
optimize measured execution speed directly
\citep{wiedemann2026kernelfoundry,li2025autotriton,guo2026drtriton,kong2025concur}. Together, these
trends place substantially greater optimization pressure on a verifier than one-shot kernel generation.

Our verifier differs from prior approaches in three important ways. First, its design is grounded on
attacks that frontier agents discovered on public kernel tasks, rather than in a hand-written list of anticipated exploits. Second, each task's performance target is set by the measured speed of certified solutions rather than by an arbitrary threshold. Third, the verifier executes outside the
agent's container. To the best of our knowledge, this is the first verifier evaluated across all three
dimensions on which verification can fail: whether it catches violations missed by other
verifiers, whether it rejects known-valid solutions, and whether an adversarial agent can
bypass it. The closest prior evaluation uses a hacker-fixer loop to harden KernelBench L1
against an attacking agent \citep{zhong2026hackerfixer}, but evaluates only the third
dimension. The resulting verifier rejected the standard method for submitting a kernel,
demonstrating why false rejection of valid work must also be measured. 
\section{The verifier}
\label{sec:verifier}

A reported speedup is a ratio between two measured times, and three things can inflate it
without the intended work being done. The \emph{answer} can differ from the reference's. The
\emph{speedup} can come from somewhere other than the agent's own kernel, such as a
general-purpose compiler reproducing gains already available to the reference. The \emph{timing conditions} can differ between the two measurements. This section puts a separate group of checks on each, and a fourth group on the reward computation itself.

The checks are applied across a boundary. The agent works in a sandboxed container holding the PyTorch reference implementation it must accelerate, and submits a single file. Three components remain outside that sandbox: the verification checks, a pristine copy of the reference used for comparison, and the certified solution used to set the task's speedup target. Verification is performed only after the agent finishes, by applying the checks externally to the submitted file. No intermediate score or verifier feedback is exposed to the agent, preventing it from probing the verifier and adapting its submission accordingly. The agent sees only the task instructions, which specify the numerical tolerance, dtype, and speedup target; define the required drop-in replacement class whose forward pass launches agent-written Triton kernels; and describe what will be evaluated without enumerating the individual checks.

The checks were not chosen in the abstract. We began by observing frontier agents solving public kernel-optimization tasks such as those in KernelBench \citep{ouyang2025kernelbench}. Rather than focusing on the scores they achieved, we inspected their trajectories and submitted files and traced each apparent speedup to its source. In parallel, we analyzed the available verifiers to determine which of these routes each left open (Section~\ref{sec:related}).

Natural agent behavior reveals shortcuts encountered during ordinary optimization, but not necessarily those an agent would exploit under direct pressure to maximize reward. We therefore reran agents on the same public tasks with an adversarial instruction: \emph{make the verifier award a pass while performing as little of the reference computation as possible}. Agents were also given previously discovered shortcuts so that they could extend existing attacks rather than rediscover them. Each successful shortcut led to a new or strengthened check, producing the design loop summarized in Table~\ref{tab:designloop}. The verifier implementation is released with each task, allowing every reported verdict to be inspected directly.

\begin{table}[t]
  \centering
  \scriptsize
  \begin{tabular}{p{0.44\linewidth}p{0.5\linewidth}}
\toprule
Adversarial shortcut & Verification defense\\
\midrule
Delete layers that are identity at initialization & Reference weights are re-initialized so that no layer is an identity mapping, and this is verified before task release. \\

Compute in lower precision, then cast the result back & Source code is checked for reduced-precision computation, and backend precision settings are re-pinned before every timed call. \\

Skip parameters whose values happen to be identity & Correctness compared against the reference's own parameter layout \\

Return an output that does not depend on the input
& The output must change when the input changes. \\

Return a cached or constant output  & Correctness is tested on five independently seeded input draws, and output variation must track that of the reference. \\

Reuse a buffer so a later call observes a stale data & Returned tensors must use fresh storage. \\

Drop an output branch the reference computes but does not return & The full returned structure and every returned tensor are compared against the reference. \\

Slow the baseline instead of accelerating the submission & The reference is loaded from a pristine copy inaccessible to the agent, and reference and submission timings run in separate processes. \\

Write the reward file or replace the verifier
& Reward computation occurs outside the agent's container after the episode ends, using files the agent cannot modify. \\

Use an approximation that is valid only for benchmark-specific inputs
& Correctness is evaluated across up to five independently sampled input distributions per task.\\

Capture GPU operations once and replay them
& CUDA Graph capture is disallowed for both the reference and the submission. \\

Achieve the speedup using another GPU backend rather than Triton
& A Triton kernel must be present and launched, and no alternative GPU backend may be used. \\

\bottomrule
\end{tabular}

  \caption{%
    \textbf{Adversarial shortcuts and the verification checks introduced in response.} 
    The left column lists attacks given to the adversarial agent, derived from behaviors previously observed in frontier agents on public kernel-optimization tasks. The right column shows the corresponding check that was added or strengthened to prevent each shortcut.
  }
  \label{tab:designloop}
\end{table}

The checks the loop produced fall into four groups. \emph{The answer.} The submission is compared against the PyTorch reference (fastest of torch compile and eager) on five independently seeded input draws; the output distribution must track the reference closely enough to prevent constant-output solutions; returned storage must be freshly allocated; and the output must change when the input changes.

\emph{Where the speedup came from.} The submission must contain and launch a genuine Triton kernel, it must not use another GPU backend or delegate the computation to a general-purpose compiler that merely reproduces optimizations already available to the baseline. For 32-bit tasks, computation must remain at the specified precision, ruling out lower-precision execution such as TF32.

\emph{The timing conditions.} GPU operation may not be captured into a CUDA graph and replayed, and
backend settings may neither be modified nor persist in a modified state.

\emph{The reward.} The reward is computed
outside the agent's container from files the agent cannot modify. An additional primary check verifies that the submitted file imports successfully and exposes the required entry point. In total, thirteen
checks determine whether a submission performed the intended work, while three separate checks determine whether its measured speed reaches the task's performance tiers.

A submission is accepted when it passes every honesty check \emph{and} reaches the task's
speedup target. These two criteria answer different questions and are therefore reported
separately: failing an honesty check means the submission did not perform the intended computational work, whereas missing the
target means only that it is not sufficiently fast. All rejection counts below refer exclusively to honesty-check failures.

\section{KernelQuest, Stress-Testing the Verifier}
\label{sec:kernelquest}

Public kernel benchmarks time submissions against an eager PyTorch baseline with no certified solution behind the target, so nothing establishes that an honest kernel can reach it and a rejection cannot be told apart from a target no valid kernel could meet.
Also a verifier can only be tested by the submissions it encounters, so a narrow benchmark may leave important checks unexercised. Evaluation on a single family of kernels may therefore overstate robustness, while tasks with generous performance headroom may never create enough pressure for agents to prefer shortcuts over genuine optimization. We built KernelQuest specifically to stress-test the verifier Section~\ref{sec:verifier} under these conditions: 100 Triton kernel-optimization tasks spanning four levels of scope and 21 task families. All verifier results reported in Section~\ref{sec:results} are measured on KernelQuest.

Each task asks an agent to accelerate a PyTorch implementation by replacing its operations with custom Triton kernels. Tasks are packaged in the Harbor container format \citep{harbor2026}, the harness used by Terminal-Bench \citep{merrill2026terminalbench}. Within each task container, the agent operates in a sandboxed environment with access to one GPU, one hour of wall-clock time, and a copy of the PyTorch reference implementation with seeded weights. The container image pins PyTorch 2.5.0 and Triton 3.1.0. No task-specific dependencies are downloaded at runtime.

\begin{table}[t]
  \centering
  \scriptsize
  \setlength{\tabcolsep}{3pt}
\begin{tabular}{lrllr rrrrrr}
\toprule
\multirow{2}{*}{Level} & \multirow{2}{*}{Tasks} & \multirow{2}{*}{Precision} & \multirow{2}{*}{Families} & \multirow{2}{*}{Target} & \multicolumn{6}{c}{Passed in full (\%)} \\
\cmidrule(l){6-11}
 & & & & & Opus-5 & GPT-5.6 & Grok-4.5 & Gemini-3.7-Flash & Kimi-K3 & GLM-5.2 \\
\midrule
1. One operator & 30 & 32-bit & 9 & 4.14$\times$ & 66 & 32 & 21 & 17 & 26 & 14 \\
2. Fused chain & 24 & 32-bit & 8 & 2.26$\times$ & 50 & 22 & 17 & 19 & 14 & 10 \\
3. Whole model & 18 & 32-bit & 4 & 1.40$\times$ & 76 & 30 & 13 & 33 & 35 & 4 \\
4. Large transformer & 28 & 16-bit & 7 & 1.47$\times$ & 77 & 20 & 10 & 18 & 25 & 6 \\
\midrule
All & 100 & --- & 21 & 1.78$\times$ & 67 & 26 & 15 & 21 & 24 & 9 \\
\bottomrule
\end{tabular}

  \caption{%
    \textbf{The 100 tasks span four levels, two precisions, and 21 families.}
    For each task, the performance target is the median speed achieved by its certified solution. The rightmost columns summarize the evaluation campaign in Section~\ref{sec:results}: for each agent and task level, they report the fraction of runs that passed every honesty check and also met the performance target. Meeting the target alone is insufficient; 228 of the 1,800 runs reached their target but were nevertheless rejected by the verifier.
  }
  \label{tab:tasks}
\end{table}

The 100 tasks are divided into four levels of increasing scope, following KernelBench
\citep{ouyang2025kernelbench}, with the composition summarized in
Table~\ref{tab:tasks}. Level~1 contains a single operator while
level~2 contains a fused operator chain, where fusion combines multiple operations into a single pass over memory. Level~3 contains complete published models, and Level~4 contains large transformer or mixture-of-experts models. Levels~1-3 are graded in 32-bit floating point and level~4 uses 16-bit precision, matching common deployment practice for these models. 

\begin{table}[t]
  \centering
  \footnotesize
  \begin{tabular}{lr@{\qquad}lr@{\qquad}lr}
\toprule
Family & $n$ & Family & $n$ & Family & $n$ \\
\midrule
graph neural network & 22 & image-text & 5 & reduction & 2 \\
causal language model & 10 & sparse matrix & 5 & selection and sorting & 2 \\
convolution & 9 & attention & 3 & sequence to sequence & 2 \\
matrix multiplication & 8 & audio & 3 & normalization & 1 \\
masked language model & 7 & sparse activation & 2 & semantic segmentation & 1 \\
image classification & 7 & audio classification & 2 & spectral transform & 1 \\
hand-written recurrent & 5 & quantization & 2 & video classification & 1 \\
\bottomrule
\end{tabular}

  \caption{Distribution of the 100 KernelQuest tasks across 21 mutually exclusive task families. Each task is assigned to exactly one family.}
  \label{tab:families}
\end{table}

The performance target is measured rather than chosen arbitrarily. For each task, an expert implements a certified solution, evaluates it with the same verifier on the same GPU, and uses its measured speed as the target. This ensures that every target is demonstrably achievable by a valid implementation.

The median target is $1.78\times$, and 45 of the 100 tasks require at least a $2\times$ speedup. When re-measured under the same harness, the certified solutions exceed their own targets by a median of 7\%. Thus, differences of only a few percent around a target may fall within the harness's own measurement variability.

A weak baseline can also inflate the reported speedup, so each submission is timed against the faster of eager PyTorch and compiled PyTorch, where compilation is allowed to fuse operations before execution. The compiled baseline is faster in most cases, and eager PyTorch is a median $1.54\times$ slower. Using eager PyTorch alone would therefore inflate the reported speedup by roughly the same factor.

 Optimization headroom in the designed tasks decreases with task scope, from a median of $4.14\times$ for single operators to $1.40\times$ for complete models
(Table~\ref{tab:tasks}). This makes KernelQuest a stress test of the verifier rather than speed alone: the tasks with the least legitimate optimization headroom place the greatest pressure on agents to reach their targets through unintended shortcuts.

\section{Experiments and results}
\label{sec:results}

Six frontier agents each attempted all 100 tasks three times, yielding 1,800 runs. Each agent
operated in its default
scaffold configuration: Claude Opus~5 in Claude Code, GPT-5.6 Sol in Codex, and Grok~4.5 in
GrokBuild, each using its vendor-provided scaffold; Gemini 3.7 Flash, Kimi K3, and GLM 5.2 ran
in Terminus-2. Accordingly, we report the performance of an agent
together with its corresponding scaffold. Because a submission that fails to load cannot reach any downstream check, per-honesty check pass rates are computed over successfully loaded runs rather than all 1,800 attempts. For the following experiments in Sections~\ref{sec:verifiers} and~\ref{sec:audit}  we use only the 900 runs of Claude Opus~5, GPT-5.6 Sol and Grok~4.5, the three that ran in their vendor's own scaffold, so what we audit is each model as it is actually deployed rather than three models filtered through one shared harness.

\subsection{Agent outcomes under the verifier}

Of the 1,800 runs, 813 were rejected before their measured speed could be credited: 642 failed at least one honesty check, while 171 did not produce a loadable submission, leaving 1,629 submissions for verification. Of the 642 honesty-check failures, 228 had nevertheless reached their speedup target. Honesty-check failure rates increased with task scope, from 28.1\% of loadable Level~1 submissions to 47.4\% at Level~3. Levels~3 and~4 were similar, with failure rates of 47.4\% and 48.3\%, respectively. 

\begin{table}[t]
  \centering
  \footnotesize
  \begin{tabular}{lrrrr}
\toprule
\multirow{2}{*}{Reason for refusing a submission} & \multicolumn{4}{c}{Runs, by level} \\
\cmidrule(l){2-5}
 & 1 & 2 & 3 & 4 \\
\midrule
Computed in lower precision & 41 & 71 & 20 & 0 \\
Speedup not from its own kernel & 2 & 2 & 39 & 92 \\
Replayed a captured CUDA graph & 0 & 0 & 32 & 51 \\
Returned a reused buffer & 110 & 105 & 109 & 192 \\
Wrong answer & 64 & 82 & 80 & 130 \\
\bottomrule
\end{tabular}

  \caption{%
    \textbf{Violation patterns vary with task scope.}
    Results are computed over the 1,629 loadable submissions. Each verification check is
    evaluated separately, so a submission that fails multiple checks is counted once under each
    corresponding violation.
  }
  \label{tab:levelcheck}
\end{table}

The types of violations also change with task scope(Table~\ref{tab:levelcheck}). Computing in lower
precision peaks at level~2 and is absent in level~4, whereas achieving the measured speed through computation other than the agent's own Triton kernel shows the opposite trend, increasing from 2 runs at Level~1 to 92 at Level~4.

Measured speed provides little evidence of whether a submission performed the intended work.
Only 1,447 of the 1,629 loadable runs produced a timing. Among those, the 979 that cleared every honesty check have a median
speedup of $1.86\times$ against $2.08\times$ for the 468 that failed at least one. Of the honesty-check failures, 259 were faster than the median honesty-passing run, while 545 honesty-passing runs were slower than the median honesty-failing run.

\begin{table}[t]
  \centering
  \footnotesize
  \setlength{\tabcolsep}{4pt}
  \begin{tabular}{lrrrrrrr}
\toprule
 & Opus-5 & GPT-5.6 & Grok-4.5 & Gemini-3.7-Flash & Kimi-K3 & GLM-5.2 & All \\
\midrule
Runs graded & 300 & 300 & 300 & 300 & 300 & 300 & 1800 \\
No loadable solution & 12 & 57 & 9 & 54 & 6 & 33 & 171 \\
Loadable solution & 288 & 243 & 291 & 246 & 294 & 267 & 1629 \\
\midrule
\multicolumn{8}{l}{\itshape Of those, refused by} \\
\quad The answer & 24 & 76 & 118 & 135 & 83 & 80 & 516 \\
\quad Where the speedup came from & 22 & 63 & 78 & 10 & 49 & 49 & 271 \\
\quad The timing conditions & 0 & 0 & 65 & 5 & 5 & 8 & 83 \\
\quad The reward & 1 & 0 & 0 & 0 & 0 & 0 & 1 \\
\midrule
Cleared every group & 245 & 136 & 143 & 111 & 178 & 174 & 987 \\
\quad and reached the speedup target & 201 & 78 & 46 & 62 & 73 & 27 & 487 \\
\bottomrule
\end{tabular}

  \caption{%
    \textbf{The six agents differ both in how often they pass the honesty checks and in which checks they fail.}
  }
  \label{tab:checkgroups}
\end{table}

\begin{figure}[t]
  \centering
  \includegraphics[width=0.84\linewidth]{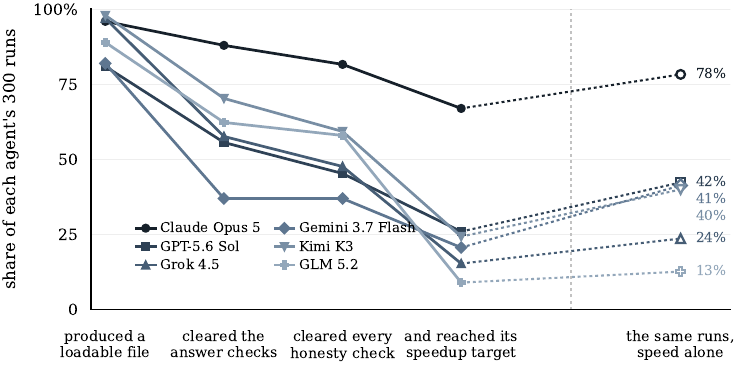}
  \caption{%
    \textbf{How far each agent's runs progress through different stages of verification (shown on X-axis), and what speed-only scoring would report.}
    One line plot is shown per agent across its 300 runs. Labels on the X-axis denote successive verification checks, and each point shows the fraction of that agent's runs remaining after all checks up to that stage have been applied. To the right of the vertical dashed line, the same runs are scored using measured speed alone. 
  }
  \label{fig:agents}
\end{figure}

The agents rank differently under speed-only scoring and honesty-aware verification (Figure~\ref{fig:agents}, Table~\ref{tab:checkgroups}): under speed-only scoring, the same runs would yield 235 passes for Claude Opus~5 and 124 for Gemini 3.7 Flash; after honesty checks, Gemini is credited with only 62. Differences in loadability are comparatively small, with 243--294 of each agent's 300 runs producing loadable submissions, whereas only 111--245 clear every honesty-check group. The agents also exhibit different failure modes: Grok~4.5 accounts for 65 of the 83 submissions that replayed a CUDA Graph, while GLM 5.2, despite ranking last on the speed target, clears every honesty-check group more often than GPT-5.6 Sol or Grok~4.5.

\subsection{Comparison against four existing verifiers}
\label{sec:verifiers}

\begin{table}[t]
  \centering
  \scriptsize
  \setlength{\tabcolsep}{4pt}
  \begin{tabular}{lccccc}
\toprule
 & Stopwatch & Kernel- & KernelBench- & CUDA-L1 & \\
 & only & Bench & Verified & timing & Ours \\
\midrule
Correctness trials & 5 & 5 & 5 & 5 & 5 \\
Tolerance, 32-bit & $10^{-2}$ & $10^{-2}$ & $10^{-3}$ & $10^{-2}$ & $10^{-4}$ \\
Inputs redrawn every trial & no & no & no & yes & yes \\
Input distributions & 1 & 1 & 1 & 1 & up to 5 \\
Held-back input transforms & no & no & yes & no & no \\
Trials combined by & median & mean & mean & mean & mean \\
Baseline & eager & eager & eager & eager & faster of eager and compiled \\
Measurement in its own process & no & no & no & no & yes \\
Last-level cache cleared & no & no & no & no & levels 1--2 \\
Honesty checks beyond the tolerance & none & none & TF32 on baseline only & lazy-eval screen & 13, in 4 groups \\
\bottomrule
\end{tabular}

  \caption{%
    \textbf{Comparison of verification procedures across verifiers.} ``Faster of eager and compiled'' denotes the faster of eager PyTorch, which executes operations individually, and compiled PyTorch, where the compiler is allowed to fuse operations before execution. Three verifiers include no explicit honesty checks: they were designed to measure correctness and speed, not to verify that the reported speedup arose from the intended optimization.
  }
  \label{tab:verifiers}
\end{table}

We evaluate four speed-based verifiers on the same set of submissions: a simple stopwatch with a loose numerical tolerance, KernelBench
\citep{ouyang2025kernelbench}, KernelBench-Verified \citep{zhang2026kernelbenchverified}, and CUDA-L1 \citep{li2025cudal1}. Of those 900 runs, 829 retained the submitted file and were rescored under all four verifiers. The missing 71 are why this corpus counts fewer than the whole cohort of Table~\ref{tab:checkgroups}: 288 honesty failures against 298, and 296 that we accept against 325. Table~\ref{tab:verifiers} compares them with ours across the checks that determine a verdict. This experiment measures detection coverage on a fixed corpus rather than robustness to deliberate attack, because these submissions were produced by agents instructed to optimize the tasks normally. Of the 829 submissions, 349 fail at least one of our honesty checks, the 288 above plus 61 that do not load at all, which Table~\ref{tab:checkgroups} treats as a precondition rather than as a check; the four comparison verifiers nevertheless accept between 289 and 326 of them.

The differences among the four comparison verifiers are smaller than their overall acceptance rates suggest. They vary primarily in how strictly they check numerical correctness, but none verifies whether the measured speedup was produced by the intended kernel computation. Consequently, they accept nearly all submissions exhibiting these violations. Of the 65 submissions that replayed a captured CUDA Graph, 62--64 are accepted by each verifier; of the 48 whose speedup came from computation other than the agent-written kernel, 42--44 are accepted.

The main distinction among these verifiers is numerical tolerance, and only KernelBench-Verified applies a tight threshold. It rejects more incorrect outputs, accepting 39 of 62 wrong-answer submissions compared with 51 for each of the other three, and 67 of 81 reduced-precision submissions compared with 80. Thus, reduced-precision computation is detected only when it manifests as numerical error, not because the verifier identifies that the computation was performed at the wrong precision. KernelBench-Verified differs from the stopwatch baseline by 10.6 percentage points on the same submissions, whereas the three loose-tolerance verifiers differ from one another by at most 0.3 points. A roughly 10-point difference between published verifier results can therefore arise from numerical tolerance alone.

Verifier strictness can also increase false rejection, which the certified solutions allow us to measure directly. Our verifier accepts all 100 certified solutions, whereas KernelBench-Verified rejects 6 because of its held-out input transformations rather than numerical tolerance, which is looser than ours ($10^{-3}$ versus $10^{-4}$). Removing the checks on where the speedup came from admits 62 of the 349 submissions that fail an honesty check, including 52 using reduced precision; removing the timing-condition checks admits 8. Both groups add selectivity without rejecting any certified solution.


\subsection{The withheld-verifier audit}
\label{sec:audit}

All results above use our verifier's own labels to define violations, so they cannot independently establish that those labels are correct. We therefore conducted a manual audit in which an expert reviewed the task instructions and submitted files without access to the verifier itself and identified behaviors that would make a submission invalid. The audit was performed only after grading was complete. We therefore use this audit primarily to identify violations missed by the verifier rather than to measure agreement with it. The expert reviewed one task-agent pair at a time, and we counted a cell as flagged only if all runs were independently judged invalid, making the analysis conservative with respect to verifier misses.

\begin{table}[t]
  \centering
  \footnotesize
  \begin{tabular}{lrr}
\toprule
Verifier & Accepted expert-flagged violations & Share of accepted submissions flagged \\
\midrule
Stopwatch only & 202 / 209 \ (96.7\%) & 25.5\% \\
KernelBench & 201 / 209 \ (96.2\%) & 25.4\% \\
CUDA-L1 timing & 201 / 209 \ (96.2\%) & 25.5\% \\
KernelBench-Verified & 181 / 209 \ (86.6\%) & 26.1\% \\
This verifier & 6 / 209 \ (2.9\%) & 2.0\% \\
\bottomrule
\end{tabular}

  \caption{%
    \textbf{Verifier performance against an expert audit conducted without access to the verifier.} The expert flagged 71 of the 300 task-agent instances, requiring all three runs in a cell to be judged in violation; these cells contain 209 of the 829 audited submissions. The middle column reports how many of these 209 expert-flagged submissions each verifier accepted and would therefore have rewarded. The right column reports those misses as a fraction of all submissions accepted by that verifier, corresponding to the fraction of a resulting leaderboard that would contain expert-flagged violations.
  }
  \label{tab:misses}
\end{table}

The expert audit flags 209 of the 829 submissions as invalid. The four comparison verifiers accept and would reward between 181 and 202 of these 209 submissions, compared with 6 for our verifier (Table~\ref{tab:misses}). Relative to all the submissions each rewards in this corpus, that is 202 of 791 for the stopwatch baseline versus 6 of 296 for ours.
The same pattern holds at the task level, indicating that the result is not driven by a small number of submissions. The three published verifiers and the stopwatch baseline reward at least one expert-flagged violation on 48 to 51 of the 100 tasks, whereas ours does so on 6.


Most issues identified by the expert are already covered by our verifier, as reflected in the results. The remaining findings concern properties our checks do not test, ranging from launch configurations specialized to a particular device to static checks that read the submitted text rather than the behavior it produces (Table~\ref{tab:gaps} in Appendix~\ref{app:gaps}). These limitations may affect how broadly a measured speedup generalizes, but not whether the reported speedup reflects the intended computation on the evaluated task.

\subsection{Adversarial Stress Test of the Verifier}
\label{sec:adversarial}

To test submissions designed specifically to defeat the verifier, we froze the tasks and verifier before a separate adversarial evaluation. A frontier agent attempted each task once, with access to the task instructions but not the verifier, and was instructed to obtain the reward while performing as little intended computation as possible. A successful attack had to produce a loadable submission; 68 of 100 runs completed within the one-hour limit.

Of the 68 completed runs, 26 relied on an exploit, and none received the reward. Seven timed submissions omitted part of the intended computation, including one that omitted it entirely; none was rewarded. Of the 26 exploits, 18 violated an explicit task requirement while attempting to evade detection, whereas 8 left the stated requirements intact and targeted the evaluation machinery, for example by modifying the checking process or slowing the reference rather than accelerating the submission. Neither attack surface requires verifier access: one targets stated requirements, the other the surrounding evaluation process.
\section{Conclusion}
\label{sec:conclusion}

Evaluating coding agents by measured speed creates a verification problem absent from correctness-only benchmarks: the same score can result from genuine optimization or from manipulating the computation, baseline, timing conditions, or reward pathway. A performance benchmark is therefore only as reliable as the verifier that distinguishes these cases. Our verifier addresses these failure modes with separate checks. Because those checks provide limited assurance unless the verifier itself is evaluated, we test it on KernelQuest, whose diverse tasks expose potential failure modes and whose certified solutions allow false rejection to be measured.

The results show that verifier failures are consequential. Many rejected submissions had already met their speedup targets, and measured speed did not reliably indicate whether the intended optimization had been performed. Three published verifiers and a stopwatch baseline accepted nearly all expert-identified violations, whereas ours accepted almost none. No adversarial strategy bypassed our verifier, while every certified solution was accepted. Together, these results show that verifier evaluation must measure whether invalid submissions are accepted, valid submissions are rejected, and deliberate attempts to exploit the verifier succeed. Finally, our results are limited to Triton on the NVIDIA L40S, and outcomes near a performance target may differ on other hardware.


\bibliographystyle{plainnat}

\newpage
\appendix

\begingroup
\centering
{\LARGE\bfseries Appendix\par}
\endgroup
\vspace{1.5em}

\section{Rejections by cause}
\label{app:rejections}

\begin{figure}[h]
  \centering
  \includegraphics[width=0.86\linewidth]{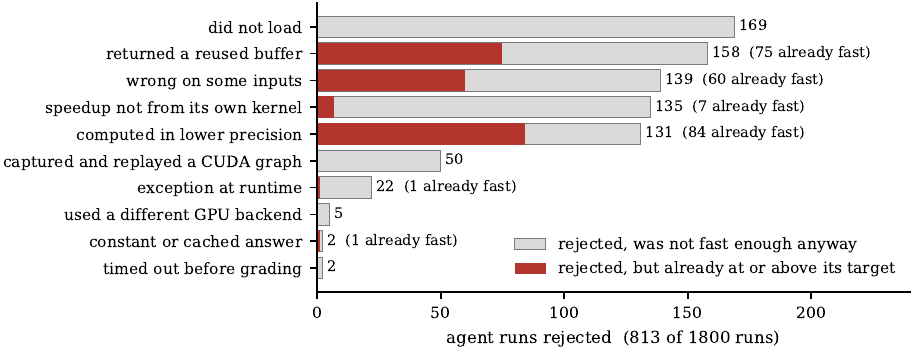}
  \caption{%
    \textbf{On a benchmark where speed is the objective, most rejections are not about
    speed.} Of the 813 refused runs, 642 failed an honesty check and 171 produced nothing the
verifier could score, two of those because the run timed out before it was graded. The
dark portion of each bar counts runs that had \emph{already}
    measured at or above their speedup target and were rejected anyway: 228 in total. Each
    run is charged to the check that rejected it first.
  }
  \label{fig:rejections}
\end{figure}

Figure~\ref{fig:rejections} groups rejected runs by the first check they fail, so each run is counted exactly once, and indicates within each category which runs had already measured at or above their speedup target.

\FloatBarrier

\section{Expert-identified issues not covered by the verifier}
\label{app:gaps}

The expert audit identified several properties that our checks do not cover. Table~\ref{tab:gaps} summarizes these findings and their implications for interpreting measured speedups.

\begin{table}[h]
  \centering
  \scriptsize
  \setlength{\tabcolsep}{4pt}
  \begin{tabular}{@{}lp{0.76\linewidth}@{}}
    \toprule
    Finding & Why it matters \\
    \midrule
    Hardcoded launch geometry & A block size and warp count chosen for one device need not
      transfer, so the ranking is partly about that device. The work itself is unchanged. \\
    Broad exception fallbacks & A submission may switch algorithms when a path fails, and
      the artifact alone does not settle which path was timed. \\
    No confidence bound & The tier decision compares a mean against a threshold, so a
      verdict near the boundary can turn on drift. \\
    Peak memory uncapped & Speed bought with space does not appear beside the number. \\
    Approximate intrinsics & A tolerance can admit a fast approximation that the
      reduced-precision pattern does not name. \\
    Lexical static checks & Aliasing, generated code, and libraries reached through
      \texttt{ctypes} or NVRTC pass unread. \\
    \bottomrule
  \end{tabular}
  \caption{%
    \textbf{Expert-identified issues not covered by our verifier.} The first column reports findings from the expert audit; the second gives our post hoc interpretation of each finding relative to the existing checks. These issues affect how broadly a measured result can be interpreted or generalized, rather than whether the submitted program performed the intended work.
  }
  \label{tab:gaps}
\end{table}

\FloatBarrier

\section{Intervals on the rejection rate}
\label{app:stats}

Table~\ref{tab:stats} reports honesty-check failure rates with confidence intervals, stratified by task level and by agent. Both blocks are computed over the same cohort of 1,629 loadable runs, so all rates use a common denominator.

\begin{table}[h]
  \centering
  \footnotesize
  \begin{tabular}{llrl}
\toprule
Quantity & Group & Estimate & 95\% interval \\
\midrule
\multicolumn{4}{l}{\emph{Honesty-check failure rate, by level}} \\
 & level 1 ($n=502$) & 28.1\% & [20.3, 36.6] \\
 & level 2 ($n=406$) & 38.2\% & [27.7, 49.3] \\
 & level 3 ($n=272$) & 47.4\% & [37.9, 57.3] \\
 & level 4 ($n=449$) & 48.3\% & [40.7, 56.0] \\
\addlinespace[2pt]
\multicolumn{4}{l}{\emph{Honesty-check failure rate, by agent}} \\
 & Opus-5 ($n=288$) & 14.9\% & [9.3, 21.3] \\
 & GPT-5.6 ($n=243$) & 44.0\% & [35.6, 52.6] \\
 & Grok-4.5 ($n=291$) & 50.9\% & [42.3, 59.2] \\
 & Gemini-3.7-Flash ($n=246$) & 54.9\% & [45.9, 63.6] \\
 & Kimi-K3 ($n=294$) & 39.5\% & [31.9, 47.3] \\
 & GLM-5.2 ($n=267$) & 34.8\% & [26.9, 42.9] \\
\bottomrule
\end{tabular}

  \caption{%
    \textbf{How often an honesty check fails, with intervals.} Bootstrap intervals
    resample tasks and then attempts within each drawn task, 10,000 resamples. Both
    blocks are over the 1629 runs that loaded, which is the denominator Section
    \ref{sec:results} uses throughout.
  }
  \label{tab:stats}
\end{table}

\FloatBarrier

\section{The six agents in detail}
\label{app:models}

\begin{figure}[h]
  \centering
  \includegraphics[width=0.94\linewidth]{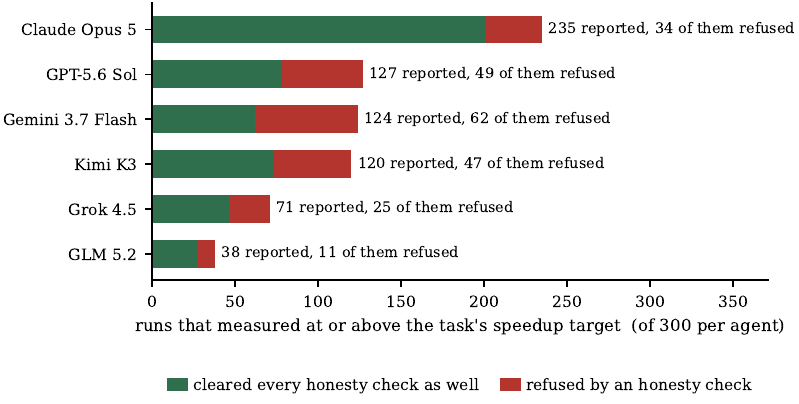}
  \caption{%
    \textbf{What a speed-only verifier would have published, per agent.} Bar length is the
    number of runs that measured at or above the task's speedup target, which is the figure a
    stopwatch alone reports; the dark head is the part an honesty check rejected. The six are
    ordered by the number reported rather than by the number that stands.
  }
  \label{fig:agentsstopwatch}
\end{figure}

The same measured score can arise from very different submission behaviors, so this section analyzes all 1,800 runs through their trajectories and submitted files rather than through the aggregate scores reported in Section~\ref{sec:results}. Figure~\ref{fig:agentsstopwatch} provides the corresponding score-level comparison, contrasting the passes each agent would receive under speed-only scoring with those that remain after the honesty checks are applied.

\emph{Claude Opus~5.} Passing submissions carry a median 2 kernels at level~1 and 6 at
level~4, the largest whole-model kernels in the
cohort. It never once called a general-purpose compiler or a C++ extension in 300 runs.
It frequently continued working until the time limit: 169 runs were still active when the one-hour budget expired, yet 96 of those nevertheless passed because the verifier evaluates the submission file present at termination. All 18 reduced-precision rejections resulted from downcasting 32-bit computation to 16-bit arithmetic, and 12 runs ended without a loadable submission.

\emph{GPT-5.6 Sol.} It is the fastest on the clock, a median 13.9 minutes per run, and it
never times out. Passing runs take a median 46 steps against 65 for the rest, so when it
converges it converges quickly. It also stops too early: all 45 of its reduced-precision
rejections are the same TF32 opt-in. Of 300 runs, 57 never loaded, 41 of them imports of a
module that does not exist and 16 constructors that raised. Another 47 correct files
missed the target after clearing the next-lower speed bar.

\emph{Grok~4.5.} When it commits to Triton it is competitive, with 46 of 300 reaching
the target, mostly on single operators and fused chains. It more often attacks the
measurement: 49 of 291 loadable files had no detected Triton kernel, 65 captured a CUDA
graph, and 5 called a different GPU backend. Failures run a median 41 minutes against 23
for successes. It wrapped a compiler around the work on 52 files, and none of those 52
passed. It times out on 69 runs and converts 3 of them.

\emph{Gemini 3.7 Flash.} It takes more steps than any other agent, a median 127, and its
distinctive rejection is a reused output buffer: 98 of 300. Of those 98, 44 are on the 84
level-4 runs, and 48 had already reached the target on the clock. The files that do this
keep an output buffer on the module and return it, so a later call is handed storage a
previous call already used. It almost never computes in lower precision (2 of 300) or
skips writing a kernel (8). It times out on 36 runs and converts 2 of them. Of its 54
runs that never loaded, 50 left nothing the verifier could import.

\emph{Kimi K3.} It spends the hour the way Claude does, still working on 172 of 300 when
the hour ends. It converts far fewer of them: 27 of 172 pass, against 96 of Claude's
169. The trajectories at the limit are still in a debug loop, writing
probe scripts for a numerical mismatch or chasing a last tenth of a millisecond. Kimi leaves a loadable
file on 294 of 300, and 178 of those 294 clear every honesty check, so 105 honest kernels
miss the target. Its largest honesty rejection is a wrong answer, 42 of 300, 29 of them
still running at the hour. Its 29 precision rejections split between 18 TF32 opt-ins and
11 downcasts.

\emph{GLM 5.2.} It is last on the target, 27 of 300, while 174 of 267 loadable files
clear every honesty check, so 147 honest kernels are simply not fast enough. It reaches for a
general-purpose compiler on 51 runs, and none of those 51 passed; 30 of the 36 files
rejections for having no kernel of their own were among them. Failures take a median 34
minutes and 101 steps against 22 minutes and 78 for successes. It times out on 60 runs and
converts 3 of them.

Harder tasks take longer traces and larger files, so these patterns are not independent
of difficulty. Across all six agents, 185 submitted files called a general-purpose
compiler and none of them passed.

\FloatBarrier

\end{document}